# FROM SU(2) GAUGE THEORY TO QUBITS ON THE FUZZY SPHERE

PAOLA ZIZZI

*Department of Brain and Behavioral Sciences, University of Pavia,*
*Piazza Botta, 11, 27100 Pavia, Italy*
paola.zizzi@unipv.it

ELIANO PESSA

*Department of Brain and Behavioral Sciences, University of Pavia,*
*Piazza Botta, 11, 27100 Pavia, Italy*
eliano.pessa@unipv.it

**Abstract**

We consider a classical pure SU(2) gauge theory, and make an ansatz, which separates the space-temporal degrees of freedom from the internal ones. This ansatz is gauge-invariant but not Lorentz invariant. In a limit case of the ansatz, obtained through a contraction map, and corresponding to a vacuum solution, the SU(2) gauge field reduces to an operator, which is the product of the generator of a global U(1) group times a Pauli matrix.

We give a geometrical interpretation of the ansatz and of the contraction map in the framework of principal fiber bundles.

Then, we identify the internal degrees of freedom of the gauge field with the non-commutative coordinates of the fuzzy sphere in the fundamental representation and obtain a one qubit state.

# 1. Introduction

The SU(2) gauge theory was the first non-abelian generalization of the U(1) gauge theory of electromagnetism. It was introduced by Yang and Mills in 1954 [1] in order to extend the SU(2) global invariance of isotopic symmetry to a local SU(2) invariance. This requires the introduction of three vector fields, one for each generator of SU(2). These non-abelian gauge fields transform according to the adjoint representation of SU(2), and must be massless, since a mass term explicitly included in the lagrangian would spoil gauge invariance. Even pure SU(2) gauge theory is highly nonlinear, and the lagrangian contains self-interaction terms. The classical solutions of the field equations of pure SU(2) gauge theory have been extensively studied by a number of authors (for a review see Ref. 2 for example). Because of the fact that the three vector fields should be associated with massless gauge bosons, pure SU(2) gauge theory was not considered as a theory of physical interest in itself. However, SU(2) gauge theory was exploited in theories where the Higgs mechanism gives masses to the gauge bosons. The first of such theories was the Glashow-Salam-Weinberg model SU(2)x U(1) of electro-weak interactions [3-5]. The second one was the "standard model" SU(3)x SU(2)xU(1) unifying strong, weak and electromagnetic interactions (for a review see, e.g., Ref. 6).

Nevertheless, we believe that a pure SU(2) gauge theory can play a very important role in understanding how a gauge field theory and quantum mechanics are related to each other. The common opinion is that quantum field theory is just quantum mechanics plus special relativity. We don't disagree completely with that, but we believe that this is not the whole story. In fact, in this paper we will show that, at least in the case of SU(2), the quantum mechanics of spin ½ can be obtained from a classical SU(2) gauge theory through a reduction mechanism.

The first stage of this mechanism was to choose an ansatz for the gauge field, which separate the infinite space-time degrees of freedom from the internal ones.

In this first step, gauge invariance is preserved, but the Lorentz symmetry is broken. Such a breaking is convenient because we want to get, as a result, quantum mechanics, which is not Lorentz invariant.

In a second step, we looked for a particular limit of the space-time dependent part of the ansatz, which leads to a vacuum solution of the field equations. In this limit, the full SU(2) gauge theory reduces to a quantum mechanical theory. This suggests that such a field theory, despite being classical, has a hidden quantum nature.

The geometrical description of such a reduction mechanism is given in terms of a local section of the principal fiber bundle, which becomes constant due to a contraction map related to the ansatz.

Then, we adopted an algebraic approach and identified the internal degrees of freedom with the non-commutative coordinates of a fuzzy sphere [7]. The internal degrees of freedom describe a one qubit state on the fuzzy sphere in the 2-dimensional representation, which has two elementary cells. Each cell encodes one bit with a given probability.

The paper is organized as follows.

In Sect. 2, we made an ansatz for the SU(2) gauge field, in terms of the product of the exponential of a U(1) gauge field times a Pauli matrix. This breaks Lorentz invariance, but not gauge invariance.

In the limit case where the U(1) gauge field tends to zero, the ansatz describes a new vacuum solution.

In Sect. 3, we considered the lagrangian density and the field equations written in terms of the ansatz, and we found an effective tensor mass which however vanishes in the vacuum.

In Sect. 4, we considered the principal fiber bundle for the SU(2) gauge theory, and made a particular choice for the U(1) gauge field in the ansatz, in such a way that the centre of an open ball in the base space topology is an attractive fixed point. The U(1) gauge field is then a contraction map in the basin of the attractor.

In Sect. 5, we showed that the contraction map acts on the open covering of the base space in such a way that the local sections become constant, and the principal fiber bundle becomes trivial. The principal connection vanishes, and the SU(2) gauge field reduces to the generator of a global U(1) times a Pauli matrix.

In Sect. 6, we expressed the inner degrees of freedom of the gauge field, resulting from the contraction map, as non-commutative coordinates on the fuzzy sphere. By introducing cells on the fuzzy sphere we found that in the case of the fundamental representation of SU(2) the fuzzy sphere describes a qubit, namely it has two cells, each one encoding one bit with a given probability. We give a geometrical picture of the quantum superposition of two cells.

Sect. 7 is devoted to the conclusions.

## 2. The Ansatz

We consider the SU(2) vector fields $A_\mu^a(x)$ ($\mu = 0,1,2,3$; $a = 1,2,3$) and make the following ansatz:

$$\tilde{A}_\mu^a(x) = e^{-i\lambda_\mu(x)}\sigma^a \tag{2.1}$$

where $\lambda_\mu(x)$ is a U(1) gauge field and the $\sigma^a$ are the Pauli matrices, which satisfy the commutation relations: $[\sigma^a, \sigma^b] = 2i\varepsilon_{abc}\sigma^c$.

The ansatz (2.1) explicitly breaks Lorentz invariance.

In (2.1) each SU(2) vector field is split into a U(1) vector field and a generator of global SU(2) in the fundamental representation, namely, a Pauli matrix. Therefore, this ansatz grasps an hidden quantum feature (spin ½) of a classical non-abelian gauge field.

Also, there are two symmetry breakings: the breaking of Lorentz symmetry, and the breaking of local SU(2).

The breaking of Lorentz symmetry is necessary to recover, in a limit we will discuss below, non-relativistic Quantum Mechanics of spin ½. The breaking of local SU(2) symmetry is realized in a different way from the usual one, which leads to the Kibble-Higgs mechanism, that is, we are not left with the coset $\frac{SU(2)}{U(1)}$, instead we have:

$$SU(2)_{local} \to U(1)_{local} \times SU(2)_{global} \qquad (2.2)$$

The gauge potentials of SU(2) are given in the matrix notation of the vector fields:

$$A_\mu(x) = \frac{1}{2i} A_\mu^a \sigma^a \qquad (2.3)$$

By inserting the ansatz (2.1) in (2.3) we get:

$$\widetilde{A}_\mu(x) = \frac{1}{2i} e^{-i\lambda_\mu(x)} \mathrm{I} \qquad (2.4)$$

where I is the $2 \times 2$ identity matrix.

The commutator for the gauge potentials (2.4) vanishes:

$$[\widetilde{A}_\mu, \widetilde{A}_\nu] = 0 \qquad (2.5)$$

Obviously, the local gauge symmetry which is left is abelian. However the fields in (2.1) keep their non-abelian character because of the presence of Pauli matrices. The abelian gauge theory cannot be identified with electrodynamics in absence of sources. In fact, the usual self-interaction of Yang-Mills theories persists in this case.

It should be stressed that in the original SU(2) theory, the non-abelian character is held by both the vector fields $A_\mu^a(x)$ and the gauge potentials $A_\mu(x)$ in (2.3) while in our case this is not true. In fact, as we have seen, the gauge fields (2.1) are non-abelian, while the gauge potentials (2.4) are abelian. However, this should not be too surprising, as the gauge potential are the basic ingredients for the gauge potential 1-form: $A = A_\mu dx^\mu$, which concerns the local gauge structure. Therefore, in our case, the gauge potential 1-form should concern only the local U(1) gauge symmetry which is left from the ansatz, (2.1) and has the expression:

$$\tilde{A} = e^{-i\lambda_\mu} dx^\mu. \tag{2.6}$$

In the ansatz (2.1) we will consider, in particular, the limit case:

$$\lambda_\mu(x) \to 0 \tag{2.7}$$

In this limit one gets:

$$\tilde{A}_\mu^a(x) \to \sigma^a. \tag{2.8}$$

In a sense, the SU(2) gauge theory reduces to the quantum mechanics of spin ½.

Let us consider the SU(2) gauge transformations performed on the general gauge potential (2.3), i.e., in absence of the ansatz:

$$A_\mu \xrightarrow{U} A_\mu' = UA_\mu U^{-1} - \frac{i}{g} U\partial_\mu U^{-1} \tag{2.9}$$

where $g$ is the gauge coupling constant, $U$ is given by:

$$U = \exp(i\rho^a(x)\sigma^a/2) \tag{2.10}$$

and $\rho^a(x)$ are three arbitrary real functions.

The gauge potential (2.4) transforms under (2.9) as:

$$e^{-i\lambda_\mu} \xrightarrow{U} e^{-i\lambda_\mu'} = e^{-i\lambda_\mu} - \frac{i}{g} U\partial_\mu U^{-1} \tag{2.11}$$

In the limit case (2.7) the transformations (2.11) become:

$$e^{-i\lambda_\mu} \xrightarrow{U} e^{-i\lambda_\mu'} = 1 - \frac{i}{g} U\partial_\mu U^{-1} \tag{2.12}$$

Eq. (2.12) can be transformed into a pure gauge by a suitable choice of the arbitrary functions $\rho^a(x)$. This means that in the limit case the ansatz (2.1) describes a vacuum solution.

In the original SU(2) theory invariant under Lorentz transformation, and the vacuum solution did correspond to $A_\mu = 0$. In presence of the ansatz, which breaks Lorentz invariance, there is, in the limit case, a new vacuum solution, corresponding to:

$$\tilde{A}_\mu = I_\mu. \tag{2.13}$$

where $I_\mu$ stands for a constant 4-vector with all components equal to 1.

## 3. The Model

In this Section, we introduce some notations, give the lagrangian density, and the field equations.

### 3. 1 Notations and conventions

Let us consider the SU(2) gauge potential in the notation (2.3).
In this notation, the expressions for the tensor field, for the covariant derivative, for the lagrangian density, and for the field equations are, respectively, given by:

$$F_{\mu\nu} = \partial_\mu A_\nu - \partial_\nu A_\mu - ig[A_\mu, A_\nu] \tag{3.1}$$

$$D_\mu = \partial_\mu - igA_\mu \tag{3.2}$$

$$L = -\frac{1}{4} F_{\mu\nu} F^{\mu\nu} \tag{3.3}$$

$$D_\mu F_{\mu\nu} = 0 \tag{3.4}$$

The expression of tensor field (3.1) in terms of the ansatz $\tilde{A}_\mu$ in (2.4) is:

$$\tilde{F}_{\mu\nu} = \partial_\mu \tilde{A}_\nu - \partial_\nu \tilde{A}_\mu \tag{3.5}$$

as a consequence of (2.5).
By using the ansatz (2.4), the expressions of the lagrangian density (3.3) andof the field equations (3.4) take the form, respectively:

$$\tilde{L} \equiv -\frac{1}{4} \tilde{F}_{\mu\nu} \tilde{F}^{\mu\nu} \tag{3.6}$$

$$\partial_\mu \tilde{F}_{\mu\nu} = 0 \tag{3.7}$$

### 3. 2 The lagrangian density and the field equations

The explicit expression of the lagrangian density (3.6) in terms of the field $\tilde{A}_\mu$ is:

$$L = -\frac{1}{4}\left[(\partial_\mu \tilde{A}_\nu)^2 + (\partial_\nu \tilde{A}_\mu)^2 - 2\partial_\mu \tilde{A}_\nu \partial_\nu \tilde{A}_\mu\right] \tag{3.8}$$

The lagrangian density (3.8) formally looks like that of U(1) electrodynamics. However, the fields $\tilde{A}_\mu$ have an expression which is not Lorentz-covariant, then the gauge boson has a mass which is not necessarily zero, but depends on the reference system.
This can be seen once the explicit expression of the ansatz (3.4) is inserted in the third term of (3.8):

$$\tilde{L} = -\frac{1}{4}\left[(\partial_\mu \tilde{A}_\nu)^2 + (\partial_\nu \tilde{A}_\mu)^2\right] - \frac{1}{2} m^2 \tilde{A}_\nu \tilde{A}^\nu \tag{3.9}$$

where:
$$m^2 = m_{\mu\nu} m^{\mu\nu}; \qquad m_{\mu\nu} = \partial_\mu \lambda_\nu. \qquad (3.10)$$

The tensor $m_{\mu\nu}$ in (3.10) represents a sort of effective tensor mass, likewise the one acquired by electrons and holes in crystals [7].

The Euler-Lagrangian equation of motion:
$$-\partial_\mu \left( \frac{\partial \widetilde{L}}{\partial(\partial_\mu \widetilde{A}_\nu)} \right) + \frac{\partial \widetilde{L}}{\partial \widetilde{A}_\nu} = 0 \qquad (3.11)$$

gives:
$$(\Box - m^2) \widetilde{A}_\nu = 0 \qquad (3.12)$$

Eq. (3.12), which describes an effective tensor mass $m_{\mu\nu}$ given in (3.10), is formally equivalent to the Proca equation [8]. However, there are some remarkable differences, one of which is that in the Proca equation the mass term is added by hand, while in Eq. (3.12) the latter arises from the choice of the ansatz. Moreover, the Proca equation is Lorentz-covariant, but not manifestly gauge invariant, while Eq. (3.12) is gauge invariant but not Lorentz-covariant.

The Noether current $J_\nu \equiv \frac{\partial L}{\partial A_\nu}$ is, in terms of the ansatz (3.4)

$$\widetilde{j}_\nu = \frac{\partial \widetilde{L}}{\partial \widetilde{A}_\nu} = -m^2 \widetilde{A}_\nu \qquad (3.13)$$

The conservation condition $\partial_\nu \widetilde{j}_\nu = 0$ implies:
$$(\partial_\nu m^2) \widetilde{A}_\nu + m^2 \partial_\nu \widetilde{A}_\nu = 0 \qquad (3.14)$$

In the Lorentz gauge the second term in (3.14) vanishes, and we get:
$$(\partial_\nu m^2) \widetilde{A}_\nu = 0 \qquad (3.15)$$

Due to the dependence of the mass $m$ on the field $\lambda_\mu$, Eq. (3.15) is in fact a constraint on the space-time dependence of this field. Such a constraint can be written explicitly as:
$$e^{-i\lambda_\nu} (\partial_\rho \lambda^\sigma) \partial_\nu (\partial_\rho \lambda^\sigma) = 0 \qquad (3.16)$$

Moreover, in the vacuum ($\lambda_\mu = 0$, $\widetilde{A}_\mu = I_\mu$), Eq. (3.15) becomes:
$$m_{\rho\sigma} \partial_\nu m^{\rho\sigma} = 0 \qquad (3.17)$$

which implies:
$$m_{\mu\nu} \equiv \partial_\mu \lambda_\nu = 0 \qquad (3.18)$$

In this way we loose the $\lambda_\mu$ space-time dependence, and then, because of the form of the ansatz, the $U(1)$ gauge field reduces to a global U(1) operator. This choice is consistent with Eq. (2.13).

From the field equations (3.12) it follows that the condition for a vanishing value of the mass, is:

$$\Box \widetilde{A}_\nu = 0 \qquad (3.19)$$

which are the analogous of the Maxwell field equations in the vacuum, that is, in absence of an external source.

By analogy, we can say that our dynamical mass plays the role of a inner source.

In terms of $\lambda_\mu$, Eq. (3.19) becomes $\Box \lambda_\nu = 0$.

Notice that, despite the absence of a mass term, Eq. (3.19) is not Lorentz covariant owing to the particular form of the ansatz $\widetilde{A}_\mu$.

## 4. The Contraction Map

The SU(2) gauge theory under consideration can be geometrically described in terms of a principal fiber bundle (for a review on principal fiber bundles see, for example, Ref. 9).

A principal fiber bundle is denoted as $(P, \pi, B, G)$, where $P$ is the total space, $B$ is the base space (in our case $R^4$), $G$ (in our case SU(2)) is the structure group, which is homeomorphic to the fiber space $F$, and $\pi$ is the canonical projection:

$$\pi : P \to R^4 \qquad (4.1)$$

The base space $R^4$ is equipped with the Euclidean metric $d$:

$$d(x', x) = |x' - x| \qquad (4.2)$$

where $x$ and $x'$ are two points of $R^4$ and must be intended as $x \equiv \{x_\mu\}$, $x' \equiv \{x_\mu'\}$ ($\mu = 1, 2, 3, 4$).

The complete metric space $(R^4, d)$ has an induced topology which is that of the open balls with rational radii $r_n = \dfrac{1}{n}$, with $n$ a positive integer.

The open ball of rational radius $r_n$, centred at $x^*$ is:

$$B_{r_n}(x^*) = \{x \in R^4 \mid d(x^*, x) < r_n\} \qquad (4.3)$$

The set of open balls $B_{r_n}(x^*)$ is an open covering of $R^4$ and forms a local basis for the topology.

Now, let us consider again the ansatz (2.1), and make the following natural choice for $\lambda_\mu(x)$:

$$\lambda(x) = x^* e^{i\frac{|x^*-x|}{n}} \tag{4.4}$$

where $\lambda$ in (4.4) must be intended as $\lambda \equiv \{\lambda_\mu\}$ $(\mu = 1,2,3,4)$.

The point $x^*$ is a fixed point for $\lambda(x)$ as it holds:

$$\lambda(x^*) = x^* \tag{4.5}$$

It is easy to check that $\lambda(x)$ continuously approaches $x^*$ for large values of $n$ (i.e., for smaller radius of the ball):

$$\lim_{n\to\infty} \lambda(x) = x^* \tag{4.6}$$

The fixed point $x^*$ is an *attractive* fixed point for $\lambda(x)$, as it holds:

$$|\lambda'(x^*)| < 1 \tag{4.7}$$

The point $x^*$ is then a particular kind of attractor for the dynamical system described by this theory.

Furthermore, it holds:

$$|\lambda'(x)| < 1 \tag{4.8}$$

for all $x \in B_{r_n}(x^*)$, which is equivalent to say that $\lambda(x)$ is a contraction mapping in the attraction basin of $x^*$, that is, it satisfies the Lipschitz condition [10].

Then, it holds:

$$d(\lambda(x), \lambda(x')) \leq q\, d(x, x') \tag{4.9}$$

with $q \in (0,1)$ for every $x, x' \in B_{r_n}(x^*)$.

## 5. Global Trivialization

Let us denote the fiber over the attractive fixed point $x^*$ by:

$$F_{x^*} \equiv \pi^{-1}(x^*) \tag{5.1}$$

In this Section, we will show that, due to the contraction mapping, all fibers $F_x \equiv \pi^{-1}(x)$ coincide with (5.1) and with the abstract fiber $F \cong SU(2)$ for every $x \in R^4$, giving rise to the trivial bundle $\pi: R^4 \times SU(2) \to R^4$.

Then, the SU(2) principal connection vanishes and the gauge field reduces to the generator of a global U(1) group times a Pauli matrix.

Let us consider the principal fiber bundle (4.1) with a local trivialization $\{\varphi_i, U_i\}$, where $\varphi_i$ is the diffeomorphism:

$$\varphi_i : \pi^{-1}(U_i) \to U_i \times SU(2) \tag{5.2}$$

and the open neighbourhood $U_i$ is the open ball (4.3), which can be expressed as:

$$U_i(x) = \left\{ x \,\middle|\, |x^* - x| < \frac{1}{n} \right\} \tag{5.3}$$

In (5.2) the map $\varphi_i$ is defined as:

$$\varphi_i(\pi^{-1}(x)) = (x, g) \tag{5.4}$$

for every $x \in U_i$ and $g \in SU(2)$.

The canonical local section associated with the local trivialization $\{\varphi_i, U_i\}$ is defined as:

$$s_i(x) \equiv \varphi_i^{-1}(x, e) \tag{5.5}$$

where $e$ is the identity element of SU(2), and it holds:

$$\pi(s_i(x)) = x, \qquad \forall x \in U_i \tag{5.6}$$

Now, let us express the open neighbourhood $U_i$ in (5.3) in terms of the contraction map (4.4):

$$U_i(\lambda(x)) = \left\{ \lambda(x) \,\middle|\, \left| -i \ln\left(\frac{\lambda(x)}{x^*}\right) \right| < \frac{1}{n^2} \right\} \tag{5.7}$$

For $n \to \infty$ we have $\lambda(x) \to x^*$, and the open neighbourhood $U_i$ becomes the singlet:

$$U_i = \{\|x^*\|\} \tag{5.8}$$

Then, because of the contraction map, all fibers over $x \in U_i$ become:

$$\pi^{-1}(x) \equiv \pi^{-1}(x^*) \tag{5.9}$$

The local trivialization (5.4) becomes:

$$\varphi_i(\pi^{-1}(x^*)) = (x^*, g) \tag{5.10}$$

and the local section (5.5) becomes a constant section:

$$s_i(x) \equiv \varphi_i^{-1}(x^*, e) = s_i(x^*) \equiv \bar{s}_i \tag{5.11}$$

The most natural choice of an atlas, in this case, is to take all the local trivialization charts of the same kind of $\{\varphi_i, U_i\}$. Let us consider for example a second chart $\{\varphi_j, U_j\}$, where $U_j$ is the open ball centred at $x^{*'}$, which can be rewritten in terms of a contraction map:

$$\lambda(x)' = x^{*'} e^{i \frac{|x^{*'} - x|}{n}} \tag{5.12}$$

and $\varphi_j$ is a local trivialization of the same kind of (5.2).

The canonical local section $s_j$ associated with the local trivialization $\{\varphi_j, U_j\}$ is defined in the same way as (6.5). For $n \to \infty$ we have $\lambda(x) \to x^{*'}$, and the open neighbourhood $U_j$ becomes the singlet:

$$U_j = \{\|x^{*'}\|\} \tag{5.13}$$

Then, because of the contraction map, all fibers over $x \in U_j$ become:

$$\pi^{-1}(x) \equiv \pi^{-1}(x^{*'}) \tag{5.14}$$

The local trivialization becomes:

$$\varphi_j(\pi^{-1}(x^{*'})) = (x^{*'}, g) \tag{5.15}$$

and the local section $s_j$ becomes a constant section:

$$s_j(x) \equiv \varphi_j^{-1}(x^{*'}, e) = s_j(x^{*'}) \equiv \bar{s}_j \tag{5.16}$$

So that we have:

$$s_i(x) = \begin{cases} s_i(x^*) & \text{for } x = x^* \\ 0 & \text{for } x \neq x^* \end{cases} \quad \text{for} \quad U_i = \{\|x^*\|\} \tag{5.17}$$

$$s_j(x) = \begin{cases} s_j(x^{*'}) & \text{for } x = x^{*'} \\ 0 & \text{for } x \neq x^{*'} \end{cases} \quad \text{for} \quad U_j = \{\|x^{*'}\|\} \tag{5.18}$$

We recall that the general relation between two local sections $s_i$ and $s_j$, canonically associated respectively with the local trivializations $\varphi_i$ and $\varphi_j$, is:

$$, s_j(x) = s_i(x) t_{ij}(x), \qquad \forall x \in U_i \cap U_j \tag{5.19}$$

where $t_{ij}(x)$ are the transition functions, which are defined by:

$$t_{ij}(x) = \varphi_i(x) \circ \varphi_j^{-1}(x) \tag{5.20}$$

Let us consider the two possible cases:

(i) $x^* \in U_i \cap U_j$

(ii) $x^{*'} \in U_i \cap U_j$

Let us consider first case (i). From (6.19) we have:

$$s_j(x^*) = s_i(x^*) t_{ij}(x^*) \tag{5.21}$$

with:

$$s_j(x^*) = 0 \tag{5.22}$$

because of (5.18).

In the same way, in case (ii) we get:
$$s_j(x^{*'}) = s_i(x^{*'})t_{ij}(x^{*'}) \tag{5.23}$$
with:
$$s_i(x^{*'}) = 0 \tag{5.24}$$
because of (5.17).
From the above we get:
$$s_j(x^{*'}) = s_i(x^{*})t_{ij}(x^{*}) \tag{5.25}$$
which is consistent only for $x^{*} = x^{*'}$ and $i = j$, leading to the identity:
$$s_i(x^{*}) = s_i(x^{*}) \tag{5.26}$$
as it is $t_{ii}(x^{*}) = 1$.

This means that if we take an atlas whose local trivialization charts are associated with the same contraction mapping, the contraction point is unique. We discard the situation where some open neighbourhoods were contracted to their centre and other were not, as the space-time base manifold would become disconnected, and in this case the fixed point attractor $x^{*}$ would be in fact a singularity.

The principal connection is defined as:
$$A_i \equiv s_i^{*}\omega \tag{5.27}$$
where $s^{*}$ is the pullback of the local section (5.5) and $\omega$ is a one-form defined on P:
$$\omega : T_p P \to \mathfrak{G} \tag{5.28}$$
In (5.28) $T_p P$ indicates the tangent space of the total space $P$ at point $p \in P$, and $\mathfrak{G}$ stands for the SU(2) algebra.

In our case, due to the contraction mapping, we have to consider the pullback of the constant section (5.11) and replace Eq. (5.27) with:
$$A_i \equiv \bar{s}_i^{*}\omega \tag{5.29}$$
We recall that the gauge potential 1-form $A_i$ is defined, for every $x \in U_i$, as:
$$A_i \equiv A_\mu dx^\mu \tag{5.30}$$
where $x_\mu$ are local coordinates in $U_i$, $A_\mu$ is the SU(2) gauge field defined as:
$$A_\mu = \frac{1}{2i} A_\mu^a \sigma^a \tag{5.31}$$
where the $\sigma^a$ are the Pauli matrices, then (5.30) can be written as:

$$A_i = \frac{1}{2i} A_\mu^a \sigma^a dx^\mu \quad (5.32)$$

By inserting the ansatz (2.1) in (5.32) we get:

$$A_i = \frac{1}{2i} e^{-i\lambda_\mu(x)} dx_\mu \quad (5.33)$$

for every $x \in U_i$.

In correspondence to the contraction point $x^*$, the connection coefficients in (5.33) become all equal to $\frac{1}{2i} e^{-ix^*}$. However, the connection $A_i$ vanishes because $dx_\mu = 0$.

The SU(2) gauge field $A_\mu^a$ reduces to an operator $A^a$ which, up to a multiplicative constant, is the product of the generator of a global U(1) group times a Pauli matrix:

$$A^a = \frac{1}{2i} e^{-ix^*} \sigma^a \quad (5.34)$$

This means that the pure SU(2) gauge field theory is reduced to a quantum mechanical theory of spin ½ with a constant U(1) "charge".

## 6. The appearance of the qubit

In the previous section, we showed that the ansatz (2.1) and the contraction map (4.4) reduced the SU(2) gauge field $A_\mu^a$ to the operator $A^a$ in (5.34). The latter is given in terms of a constant times a generator of a global SU(2) group in the fundamental representation. At this point all the dynamical aspects of the original theory disappeared, so that it will be convenient to adopt a purely algebraic approach like that of the C*-algebras [11-13].

In fact, the SU(2) algebra is a sub-algebra of the non-commutative C*- algebra of $n \times n$ complex matrices. To the latter, it is associated, by the non-commutative version of the Gelfand-Naimark theorem [14], a quantum space, which is the fuzzy sphere [15] [16].

This suggests that, it would be possible to interpret the operators (5.34) as the non-commutative coordinates of the fuzzy sphere. As we will see, this will lead to the appearance of a one qubit state.

## 6. 1 The fuzzy sphere

Let us consider the ordinary, commutative sphere $S^2$ of radius r embedded in $R^3$ : $x_1^2 + x_2^2 + x_3^2 = r^2$ \quad (6.1)

Now, let us replace the commutative Cartesian coordinates $x_i$ (i=1,2,3) of $R^3$ by the new non-commutative "coordinates" $X_i$ defined as follows:

$$x_i \to X_i = kJ_i \tag{6.2}$$

where $k$ is a parameter, called the non-commutativity parameter, and the $J_i$ are the generators of a n-dimensional irreducible representation of the Lie algebra of SU(2), satisfying the commutation relations:

$$[J_i, J_j] = 2i\varepsilon_{ijk}J_k \tag{6.3}$$

where $\varepsilon_{ijk}$ is the three-dimensional anti-symmetric tensor.

In terms of the new coordinates, Eq. (6.1) becomes:

$$X_1^2 + X_2^2 + X_3^2 = 4k^2(J_1^2 + J_2^2 + J_3^2) = 4k^2 \frac{(n^2-1)}{4} = r^2 \tag{6.4}$$

which gives:

$$k = \frac{r}{\sqrt{n^2-1}} \tag{6.5}$$

In the $n = 2$ case (the fundamental representation) the non-commutative coordinates are given in terms of the Pauli matrices:

$$x_i \to X_i = k\sigma_i \tag{6.6}$$

and Eq. (6.5) becomes:

$$k = \frac{r}{\sqrt{3}} \tag{6.7}$$

In this case, the sphere is very poorly defined as only the North and the South poles can be distinguished. However, the higher is the dimensionality $n$ of the representation, the lower is the fuzziness.

From Eq. (6.5) it follows that for $n \to \infty$, $k \to 0$ and one recovers the classical sphere $S^2$.

**6. 2 The fuzzy elementary cell**

Now, we will recall the concept of a fuzzy elementary cell, which was first introduced in [17], although it was somehow implicit in [15] through the introduction of the constant:

$$K = 4\pi kr \tag{6.8}$$

which has the dimension of a squared length.

By inserting the explicit expression of $k$ in (6.8) one gets:

$$A_{(n)} = \frac{4\pi r^2}{\sqrt{n^2 - 1}} \tag{6.9}$$

which is the quantized surface area of the fuzzy sphere.
This is the area to be shared by $n$ elementary cells.
From Eq. (6.9) it follows that for $n \to \infty$, $A_{(n)} \to 0$, and the cells reduce to points, as $k \to 0$.
Furthermore, we require that each elementary cell encodes one string of $n$ bits, the basis states of an Hilbert space $C^n$ with $n = 2^N$, where N is an integer labelling the number of qubit states. We will denote the $i^{th}$ string as $\xi_i$
The area of the $i^{th}$ elementary cell is then:

$$A_{(n)i} = \frac{4\pi r^2}{\sqrt{n^2 - 1}} p(\xi_i) \tag{6.10}$$

where $p(\xi_i)$ is the probability of finding the string $\xi_i$ in the $i^{th}$ cell.
It should be noticed that the strings $\xi_i$ are the cyclic vectors of the Hilbert space $C^{2^N}$, which can be obtained from pure states of the non-commutative C*-algebra of $n \times n$ complex matrices (with $n = 2^N$) through the Gelfand-Naimark-Segal construction.
As it must hold:

$$\sum_{i=1}^{n} p(\xi_i) = 1 \tag{6.11}$$

it follows that not all the $n$ irreducible representations are allowed, but just those with: $n = 2^N$.
The fuzzy sphere in the $n = 2^N$ representation, encoding one string of $n$ bits in each elementary cell, reminds the holographic principle [18] [19].
In the particular case $n = 2$, there are two elementary cells sharing the quantized area:

$$A_{(2)} = \frac{4\pi r^2}{\sqrt{3}} \tag{6.12}$$

The area of the two elementary cells are:

$$A_{1\,(2)} = \frac{4\pi r^2}{\sqrt{3}} p(\xi_1), \qquad A_{2\,(2)} = \frac{4\pi r^2}{\sqrt{3}} p(\xi_2) \tag{6.13}$$

where $\xi_1$, $\xi_2$ are the two cyclic vectors:

$$\xi_1 = \begin{pmatrix} 1 \\ 0 \end{pmatrix}, \quad \xi_2 = \begin{pmatrix} 0 \\ 1 \end{pmatrix} \tag{6.14}$$

### 6.3 The qubit on the fuzzy sphere

Now, let us go back to the starting point of this discussion, namely to Eq. (5.34), where the operator $A^a$ $(a = 1,2,3)$ can be viewed as a non-commutative coordinate:

$$A^a \equiv X^a = k\sigma^a \tag{6.15}$$

with:

$$k = \frac{1}{2i} e^{-ix^*} \tag{6.16}$$

By the use of the Euler formula, the operator $A^a$ in (5.34) can be rewritten as:

$$A^a = -\frac{1}{2}\left(i\cos x^* + \sin x^*\right)\sigma^a \tag{6.17}$$

Let us consider the real and imaginary parts of $A^a$:

$$\operatorname{Im} A^a = -\frac{1}{2}\cos x^* \sigma^a \tag{6.18}$$

$$\operatorname{Re} A^a = -\frac{1}{2}\sin x^* \sigma^a \tag{6.19}$$

as the components of a vector in the Argand-Gauss plane:

$$\begin{pmatrix} \sin x^* \\ \cos x^* \end{pmatrix} = \sin x^* \begin{pmatrix} 1 \\ 0 \end{pmatrix} + \cos x^* \begin{pmatrix} 0 \\ 1 \end{pmatrix} \tag{6.20}$$

which is a qubit, since it holds: $\sin^2 x^* + \cos^2 x^* = 1$.

Notice that the area of the two cells in (6.13) are equal only for $x^* = \frac{\pi}{4}$, to which it corresponds: $\sin x^* = \cos x^* = \frac{1}{\sqrt{2}}$, so that $p(\xi_1) = p(\xi_2) = \frac{1}{2}$.

In this case, the two cells in (6.13) have equal area, given by:

$$A_{1_{(2)}} = A_{2_{(2)}} \equiv A_{(2)} = \frac{2\pi r^2}{\sqrt{3}} \tag{6.21}$$

and Eq.(6.20) is the so-called cat state:

$$|Q\rangle = \frac{1}{\sqrt{2}}\left(|0\rangle + |1\rangle\right) \tag{6.22}$$

The unitary operator $U$ which interchanges the two cyclic states $\xi_1$ and $\xi_2$, namely $U(\xi_1) = \xi_2$, $U(\xi_2) = \xi_1$, is just the NOT logic gate:

$$NOT = \begin{pmatrix} 0 & 1 \\ 1 & 0 \end{pmatrix} \quad (6.23)$$

which interchanges the two bits $|0\rangle = \begin{pmatrix} 1 \\ 0 \end{pmatrix}$ and $|1\rangle = \begin{pmatrix} 0 \\ 1 \end{pmatrix}$:

$$NOT|0\rangle = |1\rangle \qquad NOT|1\rangle = |0\rangle \quad (6.24)$$

and therefore exchanges the real with the imaginary parts of $A^a$.

### 6. 4 Geometrical quantum superposition

As we have seen, the fuzzy sphere in the $n = 2$ representation, has two cells, which can be numbered in the binary base and identified with the two bits $|0\rangle$ and $|1\rangle$, each one with a given probability. In the particular case discussed above, the probabilities were taken to be equal to $\frac{1}{2}$, and the two cells had the same area. See fig. 1.

**Fig. 1**
**The fuzzy sphere with two cells.**

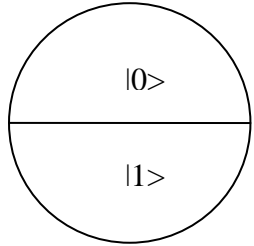

The algebraic way to get a cat state from one bit is the use of the Hadamard gate $H$:

$$H|0\rangle = \frac{1}{\sqrt{2}}(|0\rangle + |1\rangle) \qquad H|1\rangle = \frac{1}{\sqrt{2}}(|0\rangle - |1\rangle) \quad (6.25)$$

where: $H = \frac{1}{\sqrt{2}}\begin{pmatrix} 1 & 1 \\ 1 & -1 \end{pmatrix}$

We are interested in the corresponding geometrical description of the cat state on the fuzzy sphere. When the cell identified with $|0\rangle$ is rotated across the equator, the result is a cat state $\frac{1}{\sqrt{2}}(|0\rangle + |1\rangle)$ at the North pole, while the rotation of the cell identified with $|1\rangle$ gives a cat state $\frac{1}{\sqrt{2}}(|0\rangle - |1\rangle)$ at the South pole. See Fig.2.

**Fig. 2**
**Quantum superposition of the two cells.**

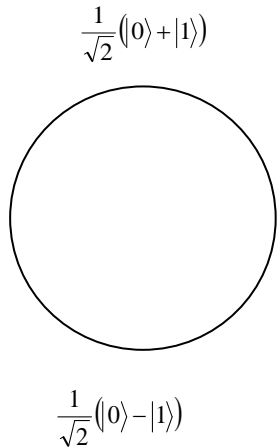

Notice that in this way we recover the "degree" of fuzziness described by Madore in [15] for $n = 2$, that is, the North and the South poles are the only points which can be distinguished. The general case with $N \geq 2$ qubits maximally entangled on the fuzzy sphere in the $n = 2^N$-dimensional representation is under study.

**6.5 The fuzzy sphere and Loop Quantum Gravity**

From Eq. (6.16), we have, for $x^* = \dfrac{\pi}{4}$:

$$k = -\frac{\sqrt{2}}{4}(1+i) \tag{6.26}$$

From Eq. (6.7), and Eq. (6.26) it follows that the radius of the fuzzy sphere is a complex number $\lambda$:

$$r \equiv \lambda = -\frac{\sqrt{6}}{4}(1+i) \tag{6.27}$$

with $|\lambda| = \dfrac{\sqrt{3}}{2}$.

In this case the area in (6.21) takes the numerical value:

$$A_{(2)} = \frac{\pi\sqrt{3}}{2} \tag{6.28}$$

In Loop Quantum Gravity (LQG) [20] [21] the discrete area spectrum [22] is:

$$A = 8\pi l_P^2 \gamma \sum_i \sqrt{j_i(j_i+1)} \tag{6.29}$$

where $l_P \approx 10^{-33} cm$ is the Planck length, the $j_i$ are the spins in the irreducible representation of SU(2) which label the spin networks' edges, and $\gamma$ is the Immirzi parameter [23].

It should be noticed that in the case of one puncture of a spin in the fundamental representation of SU(2), namely for $j = \dfrac{1}{2}$, Eq. (6.29) reduces to:

$$A = 8 l_P^2 \gamma\, A_{(2)} \tag{6.30}$$

with $A_{(2)}$ given in (6.28).

Then, the area of the elementary cell of the fuzzy sphere in the fundamental representation of SU(2) is proportional to the minimal area of the LQG spectrum for a given value of the parameter $\gamma$.

**7. Conclusions**

In this paper, we described the reduction of the pure SU(2) gauge theory down to the quantum mechanics of spin ½ in terms of an ansatz for the gauge field in the vicinity of an attractive fixed point.

There was at first a net separation (through the ansatz) between space-time and internal degrees of freedom in the SU(2) gauge field. Then, the spatio-temporal degrees of freedom were lost in one attractor by means of a contraction map. The survived internal degrees of freedom were those of a global SU(2) symmetry group. This was the procedure to get quantum mechanics from a classical gauge field theory, of course at the expenses of Lorentz invariance, which is broken by the ansatz.

At this point, it was possible to take an algebraic approach to the quantummechanics of global SU(2) by using the (non-commutative) C*-algebra formalism, and the generalization of the Gelfand duality to non-commutative (quantum) spaces. In fact, it appeared that the remaining internal degrees of freedom could be viewed as the non-commutative coordinates of a fuzzy sphere in the fundamental representation. The result was a qubit embedded in a quantum space, as the geometrical representation of the qubit state space (the Bloch sphere) is in a one-to-one correspondence with the fuzzy sphere in the fundamental representation. The "reduced field" $A^a$ became a non-commutative coordinate of the fuzzy sphere, giving rise to a one qubit state, whose state space (the Bloch sphere) is identified with its quantum embedding space [24] (the fuzzy sphere with two cells). In this way the qubit state becomes inaccessible to any external observer, or environment.

This scenario describes a quantum-computing system identified with a quantum space.

The quantum computer, the quantum space, and the quantum system are identified. This should happen for quantum gravity in particular [17], but perhaps it is a more general concept.